\documentclass[a4paper,fleqn]{cas-sc}

\usepackage{bm}
\usepackage[protrusion=true,expansion=true]{microtype}
\usepackage[normalem]{ulem}
\usepackage{soul}
\usepackage{xcolor}
\usepackage{kotex}
\usepackage{hyperref}
\usepackage{cleveref}
\usepackage{comment}

\usepackage[numbers]{natbib}
\usepackage{float}
\usepackage{placeins}

\def\tsc#1{\csdef{#1}{\textsc{\lowercase{#1}}\xspace}}

\tsc{WGM}
\tsc{QE}

\begin{document}
\let\WriteBookmarks\relax
\def\floatpagepagefraction{1}
\def\textpagefraction{.001}

\shorttitle{Breathing chimera states from purely triadic interactions}

\shortauthors{Yi et al.}

\title [mode = title]{Breathing chimera states from purely triadic interactions}

\author[1]{Sudo Yi}
\fnmark[1]
\credit{Conceptualization, Investigation, Writing - original draft \& editing}

\author[2]{Gug Young Kim}
\fnmark[1]
\credit{Conceptualization, Investigation, Writing - review \& editing}

\author[3]{Mi Jin Lee}
\cormark[1]
\ead{mijinlee@pusan.ac.kr}
\credit{Supervision, Funding acquisition, Writing - review \& editing}

\author[2]{S.-W. Son}
\cormark[1]
\ead{sonswoo@hanyang.ac.kr}
\credit{Supervision, Project administration, Funding acquisition, Writing - review \& editing}

\author[1]{B. Kahng}
\cormark[1]
\ead{bkahng@kentech.ac.kr}
\credit{Supervision, Project administration, Funding acquisition, Writing - review \& editing}

\affiliation[1]{organization={CCSS, KI for Grid Modernization, Korea Institute of Energy Technology},
            addressline={21 Kentech-gil},
            city={Naju-si},
            postcode={58330},
            state={Jeollanam-do},
            country={Korea}}
            
\affiliation[2]{organization={Department of Applied Physics, Hanyang University ERICA},
            addressline={55 Hanyangdaehak-ro},
            city={Ansan-si},
            postcode={15588},
            state={Gyeonggi-do},
            country={Korea}}

\affiliation[3]{organization={Department of  Physics, Pusan National University},
            addressline={2 Busandaehak-ro 63beon-gil},
            city={Geumjeong-gu},
            postcode={46241},
            state={Busan},
            country={Korea}}

\cortext[1]{Corresponding authors}

\fntext[1]{These authors contributed equally to this work.}
\begin{abstract}
Chimera states, characterized by the coexistence of synchronized and desynchronized dynamics in identical oscillators, are typically studied in systems with pairwise interactions. Recent studies have shown that purely higher-order interactions can support chimera states. Here, we show that breathing chimera states arise in a model governed solely by a triadic phase interaction, whose intrinsic $\pi$-symmetry permits bimodal phase distributions. We construct a bimodal Ott–Antonsen reduction that incorporates an asymmetry parameter via prescribed asymmetric initial conditions, thereby achieving an exact low-dimensional description of the macroscopic dynamics. This allows us to derive an analytic condition for the emergence of chimera states and identify a bifurcation to a breathing chimera regime characterized by persistent oscillations. Furthermore, the reduced dynamics can be expressed as a Riccati-type equation, providing a geometric interpretation of the chimera state as a closed periodic orbit in the complex plane. 
Our results establish purely triadic coupling as a minimal higher-order mechanism for chimera formation and provide a tractable framework for studying symmetry-broken collective dynamics in systems dominated by many-body interactions.\\

\end{abstract}

\begin{highlights}
\item Chimera states can emerge from purely triadic interactions without any pairwise coupling.
\item The intrinsic $\pi$-symmetry of the triadic coupling permits bimodal phase distributions.
\item A bimodal Ott–Antonsen ansatz with an asymmetry parameter yields an exact reduction.
\item Riccati-type reduced dynamics reveals breathing chimeras as closed periodic orbits in the complex plane. 
\end{highlights}



\maketitle

\section{Introduction}
The phenomenon of chimera states, where synchronized and desynchronized oscillators coexist within a population of identical oscillators, has become a central topic in the study of complex systems ~\cite{Kuramoto2002,Abrams2004,Abrams2006,Panaggio2015,Parastesh2021}. As a striking example of spontaneous symmetry breaking, chimera states demonstrate how macroscopic heterogeneity can emerge even in systems that are structurally homogeneous and composed of identical units~\cite{Kuramoto2002,Abrams2004}. 
Beyond numerous theoretical and experimental studies, chimera states have recently been observed in natural systems, most notably in synchronous firefly swarms, highlighting their broader relevance to collective dynamics~\cite{Sarfati2022}.
This counterintuitive behavior was originally motivated by biological observations such as unihemispheric sleep, observed in certain birds and marine mammals, where one cerebral hemisphere remains synchronized while the other stays desynchronized and awake~\cite{Rattenborg2000,Bohm2015}. Such phenomena raise a fundamental question: 
how does the interaction structure shape the emergence and dynamics of chimera states?

From a theoretical perspective, extensive efforts have been devoted to identifying the minimal conditions under which chimera states can emerge. Within the framework of pairwise-coupled oscillator systems, it has been revealed that phase lag and nonlocal coupling can facilitate the coexistence of coherent and incoherent subpopulations~\cite{Kuramoto2002,Abrams2004,Abrams2008,Yi2022}. In these settings, analytical approaches based on the Ott–Antonsen (OA) ansatz have played a crucial role by enabling a reduction of the high-dimensional dynamics to a tractable set of low-dimensional equations, thereby allowing systematic analysis of the existence and stability of chimera states~\cite{Abrams2008,Ott2008,Laing2009}. 

Meanwhile, many real-world systems may involve higher-order interactions, in which three or more elements interact simultaneously. Examples range from neural systems, where collective activity depends on interactions among many neurons rather than on pairwise coupling alone, to ecological and social systems governed by group-level dynamics. Recent developments in network science have emphasized that such higher-order interactions can fundamentally alter the collective behavior of dynamical systems, leading to enhanced multistability, abrupt transitions, and novel synchronization patterns~\cite{Skardal2019,Skardal2020,Millan2020,Boccaletti2023,Tanaka2011,Kundu2022,Luo2024}. However, higher-order interactions are often considered together with pairwise coupling, 
making their intrinsic dynamical effects difficult to isolate. 
Although chimera states have recently been reported in systems with purely higher-order interactions~\cite{Kar2024,Muolo2024}, those studies {employed} different many-body coupling structures. 
The resulting collective dynamics and analytical tractability can depend strongly on the specific form of the higher-order phase coupling. Here, we investigate
whether the triadic phase coupling specified below can generate chimera
states and admit an exact low-dimensional description.

In this work, we consider a minimal higher-order oscillator model that incorporates purely triadic interactions within a simple two-group structure. By explicitly excluding pairwise coupling, we isolate the intrinsic dynamical effects of higher-order interactions and demonstrate that they are sufficient to induce chimera states. 
The resulting triadic dynamics possesses an intrinsic $\pi$-symmetry, supporting bimodal phase distributions that cannot be captured by the standard OA ansatz.

To overcome this difficulty, we develop a modified OA ansatz that incorporates an asymmetry parameter to describe bimodal phase distributions. This formulation yields an exact dimensional reduction of the higher-order oscillator dynamics to a closed low-dimensional system. Building on this exact reduction, we analytically derive the criterion for the emergence of chimera states 
and show, using the exact Riccati-type dynamics and its conserved invariant, that the breathing dynamics consists of a neutral family of closed periodic orbits, in which the partially synchronized population undergoes persistent periodic oscillations.

\section{Model and Order Parameter} 
We consider a system of coupled oscillators organized into two distinct groups with a phase lag, as suggested by~\cite{Abrams2008}. Here, we introduce purely higher-order interactions instead of pairwise ones, and the resulting dynamics of each oscillator $i$ in group $\sigma$ is described by the following equation:
\begin{equation}
\frac{d\theta_i^\sigma}{dt} = \omega + \sum_{\sigma^\prime} \sum_{\sigma^{\prime\prime}} \frac{K_{\sigma \sigma^\prime\sigma^{\prime\prime}}}{N_{\sigma^\prime} N_{\sigma^{\prime\prime}}} \sum_{j=1}^{N_{\sigma^\prime}} \sum_{k=1}^{N_{\sigma^{\prime\prime}}} \sin(\theta_j^{\sigma^\prime} + \theta_k^{\sigma^{\prime\prime}} - 2\theta_i^\sigma - \alpha)~,
\label{eq:eq_of_motion}
\end{equation}
where the group indices $\sigma,\sigma^{\prime},\sigma^{\prime\prime}\in\{1,2\}$. The interaction term represents triadic coupling among oscillators, with $\sigma^\prime$ and $\sigma^{\prime\prime}$ denoting the groups from which the two interacting oscillators are selected.

\begin{figure*}
\centering
\includegraphics[width=\textwidth]{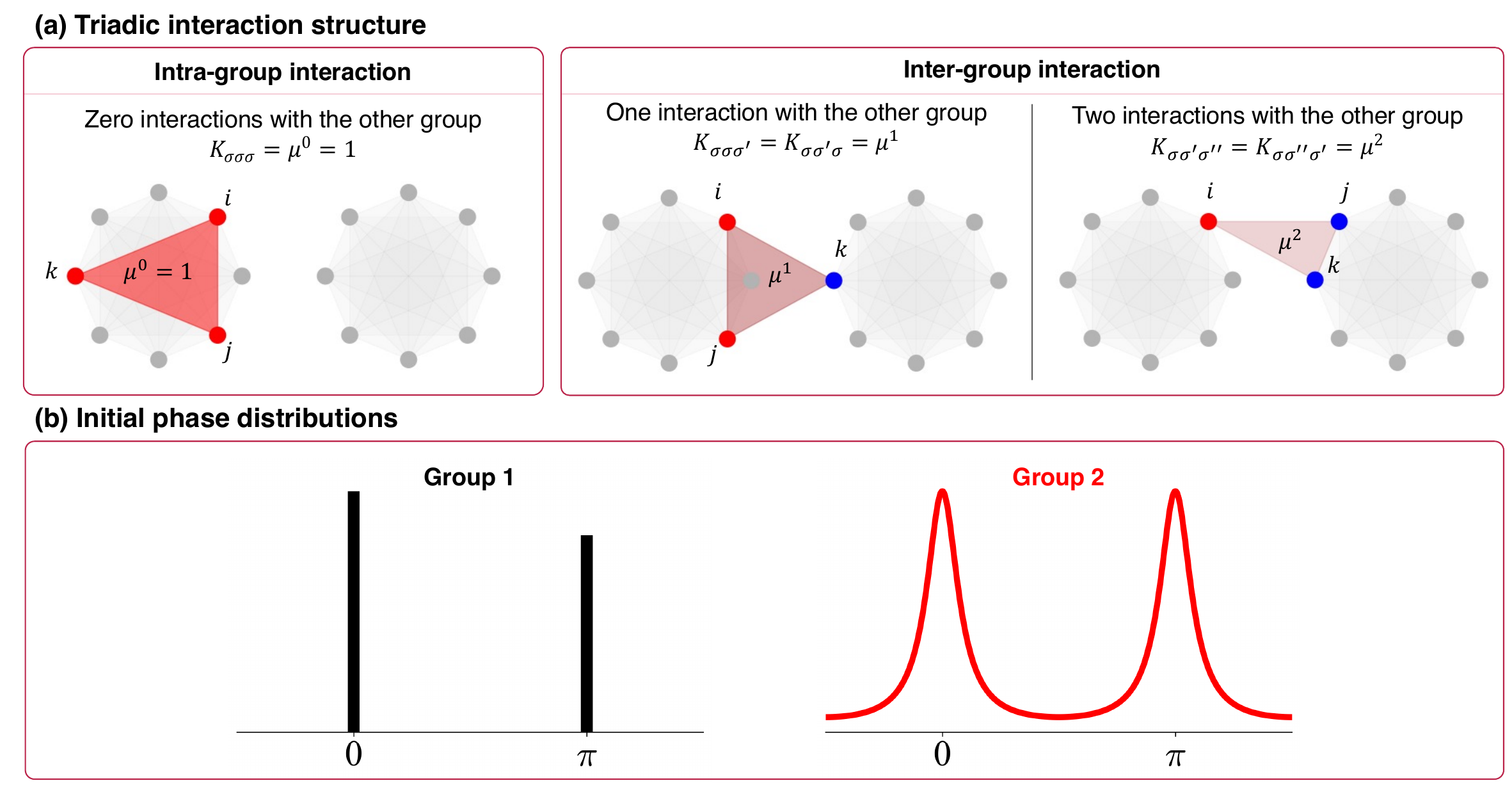}
\caption{Schematic view of the coupling strength $K_{\sigma \sigma^{\prime} \sigma^{\prime\prime}}$ and the initial phase distributions. 
(a) Intra-group coupling is assumed to be the strongest, and coupling is weakened by a term $\mu$ depending on the number of oscillators in the opposite group. 
(b) Initial phase distributions $f_{\sigma}(\theta,t=0)$ of the two groups. Group 1 is prepared in an asymmetric synchronized state with $\eta_1 \ne 0.5$ and $Q_1=1$, whereas Group 2 is initialized in  a partially synchronized state with a symmetric bimodal distribution with $\eta_2=0.5$ and $Q_2<1$.}
\label{fig:fig1}
\end{figure*}

 
Two representative forms of triadic phase coupling acting on oscillator $i$ are $\sin(\theta_j+\theta_k
-2\theta_i)$ and
$\sin(2\theta_j-\theta_k
-\theta_i)$, which have been studied in
Refs.~\cite{Skardal2019,Skardal2020}, respectively.
Here, we consider the former, in which the two source phases enter on equal footing and the coupling depends on the second harmonic of the receiving phase.
Consequently, the coupling function is invariant under
$\theta_i\rightarrow\theta_i+\pi$, which we refer to as
its intrinsic $\pi$-symmetry. This structure motivates the bimodal OA
reduction developed below.
The coefficients $K_{\sigma\sigma^\prime\sigma^{\prime\prime}}$ determine the coupling strength for each combination of these group labels. The strength of interaction decreases as the group of the receiving oscillator differs from the group(s) of the participating oscillator(s) [see Fig.~\ref{fig:fig1}(a)]. The interaction term is normalized by the number of oscillators in the respective groups, ensuring that the coupling effect does not scale trivially with system size. The presence of the phase lag parameter $\alpha$ plays a crucial role in modulating the interaction, affecting the emergence and stability of synchronization patterns, including chimera states. Here, we consider that the number of oscillators $N_\sigma=N$ for all $\sigma$'s so that the system contains a total of $2N$ oscillators.

The coupling tensor $K_{\sigma\sigma^\prime\sigma^{\prime\prime}}$ is determined by the group composition of the interacting triad. As schematically shown in Fig.~\ref{fig:fig1}(a), the coupling strength in this system is given by
\begin{equation}
K_{\sigma \sigma^{\prime} \sigma^{\prime\prime}} = \mu^{(1-\delta_{\sigma \sigma^{\prime}}) + (1-\delta_{\sigma \sigma^{\prime\prime}})}~,
\end{equation}
where $\mu$ is defined by the inter-group interaction parameter ($\mu \in [0, 1)$) and $\delta$ denotes the Kronecker delta. This formulation determines the interaction strength based on the group membership of the interacting oscillators. We set the intra-group triadic interaction strength to unity, $K_{\sigma\sigma\sigma}=1$, and assume that the coupling is weakened by a positive factor $\mu<1$ whenever an oscillator from the other group participates in the interaction. Accordingly, configurations containing one oscillator from the opposite group have strength $\mu$, while those containing two oscillators from the opposite group have strength $\mu^2$. This structure ensures that intra-group interactions remain strong while inter-group interactions are suppressed depending on the value of $\mu$, which plays a crucial role in the emergence of chimera states. 

{\it Order parameter ---}
The phase synchronization is typically quantified by the Kuramoto order parameter
\begin{equation}
   R_\sigma e^{{\rm i}\psi_\sigma}=
\frac{1}{N_\sigma}\sum_{j=1}^{N_\sigma} e^{{\rm i}\theta_j^\sigma}~,
\end{equation}
where $R_\sigma$ captures the coherence of oscillator phases in a group $\sigma$. In the present triadic interaction model, however, the intrinsic $\pi$-symmetry of the phase dynamics in Eq.~\eqref{eq:eq_of_motion} permits the formation of bimodal phase distributions, in which two clusters separated by $\pi$ coexist with $R_\sigma\approx 0$, even when the system exhibits a high coherence of phase synchronization. Therefore, in the present system with $\pi$-symmetry, it is more appropriate to employ the Daido order parameter~\cite{Daido1996} defined as 
\begin{equation}
    Q_\sigma e^{{\rm i}\chi_\sigma} =
\frac{1}{N_\sigma}\sum_{j=1}^{N_\sigma} e^{{\rm i}2\theta_j^\sigma}~,
\end{equation}
which remains sensitive to bimodal phase coherence.

In the following, we define a chimera state as a configuration in which one group exhibits a fully phase-locked bimodal distribution ($Q_1=1$), while the other remains only partially synchronized ($Q_2<1$). This distinction allows us to characterize the coexistence of coherent and incoherent dynamics within the higher-order interaction framework.

\section{Modified OA Ansatz} 

The intrinsic $\pi$-symmetry of the triadic interaction supports stable bimodal phase distributions, which cannot be described by the standard OA ansatz assuming unimodal coherence~\cite{Ott2008}. To incorporate this bimodal structure, we decompose the phase density into two anti-phase subpopulations,
\begin{equation}
f_\sigma
=
\eta_\sigma f_{\sigma,a}
+
(1-\eta_\sigma)f_{\sigma,b}~,
\label{eq:initial_density}
\end{equation}
where the adjustable value $\eta_\sigma$ is the asymmetry parameter of the distribution [Fig.~\ref{fig:fig1}(b)] and the two terms in Eq.~\eqref{eq:initial_density} satisfy the $\pi$-shift symmetry
$f_{\sigma,b}(\theta)=f_{\sigma,a}(\theta-\pi)$.
Applying the OA ansatz separately to each subpopulation allows us to rewrite the phase density by decomposing odd and even terms as
\begin{equation}
    f_{\sigma} = \frac{1}{2\pi} \left( 1+ \left[ \sum_{n=1}^{\infty}{a_{\sigma}^{2n} e^{{\rm i}2n\theta}}+(2\eta_{\sigma}-1)\sum_{n=1}^{\infty}{a_{\sigma}^{2n-1} e^{{\rm i}(2n-1)\theta}}\right] + \mathrm {c.c}.\right),
    \label{eq:f_OAform}
\end{equation}
and yields a reduced bimodal manifold parameterized by the asymmetry parameter $\eta_\sigma$ (see Supplemental Material Sec.~{S1}.B for details). Under this construction, the macroscopic dynamics remains reducible to a low-dimensional manifold described by a complex amplitude
$a_\sigma\equiv r_\sigma e^{-{\rm i}\psi_\sigma}$, 
where $r_\sigma\in[0,1]$ is the real-valued modulus of the complex amplitude and represents the degree of synchronization within each peak of the bimodal distribution.

A direct consequence of this bimodal OA formulation is that the Kuramoto and Daido order parameters become
\begin{equation}
R_\sigma =\left|2\eta_\sigma-1\right|r_\sigma~,
\qquad
Q_\sigma = r_\sigma^2~.
\label{eq:order}
\end{equation}
{See Supplemental Material Sec.~S1.B for the derivation of Eq.~\eqref{eq:order} from the modified bimodal OA ansatz.}
These expressions reveal that the first-order coherence $R_\sigma$ is controlled explicitly by the asymmetry parameter $\eta_\sigma$, whereas the second-order coherence $Q_\sigma$ depends only on the modulus of the complex amplitude $a_\sigma$, i.e., $r_\sigma$. In particular, the symmetric bimodal state $\eta_\sigma=0.5$ gives $R_\sigma=0$, even when the second-order coherence is maintained ($Q_\sigma>0$).

This framework enables an exact dimensional reduction of the microscopic dynamics into a small set of macroscopic variables. Here, the asymmetry parameter $\eta_\sigma$ determines the initial imbalance between the two anti-phase subpopulations and remains conserved throughout the dynamics, since the two subpopulations evolve independently within the reduced manifold. Consequently, $\eta_\sigma$ controls the initial asymmetry and thereby influences the emergence of chimera states.

{\it Initial conditions ---}
It is well known that, since higher-order interaction systems exhibit strong multistability, the resulting macroscopic dynamics depends sensitively on the initial condition~\cite{Skardal2019,Zhang2024}. To induce chimera states, we prepare asymmetric initial phase distributions inspired by unihemispheric sleep ~\cite{Rattenborg2000}, as illustrated in Fig.~\ref{fig:fig1}(b): one group ($\sigma=1$) is initialized in an asymmetric coherent state with $\eta_1 \neq 0.5$ and $Q_1 = 1$, while the other ($\sigma=2$) remains symmetric with $\eta_2 = 0.5$ and partial synchronization $0<Q_2 < 1$. This asymmetry provides the initial condition from which the coexistence of synchronized and partially synchronized collective states emerges.
{The specific initial conditions used in the numerical simulations are given in Supplemental Material Sec.~S1.C.}
Figure~\ref{fig:phase_distribution} shows the phase distributions of the two groups at the selected minimum of $Q_2$. Group~1 maintains a phase-locked bimodal state with $Q_1=1$, while the phases in Group~2 are spread over a broad range. The doubled-phase representation shows that Group~2 does not form distinct phase-locked clusters. Supplementary Video~S2 shows that the phase distribution of Group~2 changes periodically as $Q_2$ oscillates. Thus, a fully synchronized bimodal state in Group~1 coexists with a breathing partially synchronized state in Group~2.

\begin{figure*}[b]
\centering
\includegraphics[width=\textwidth]{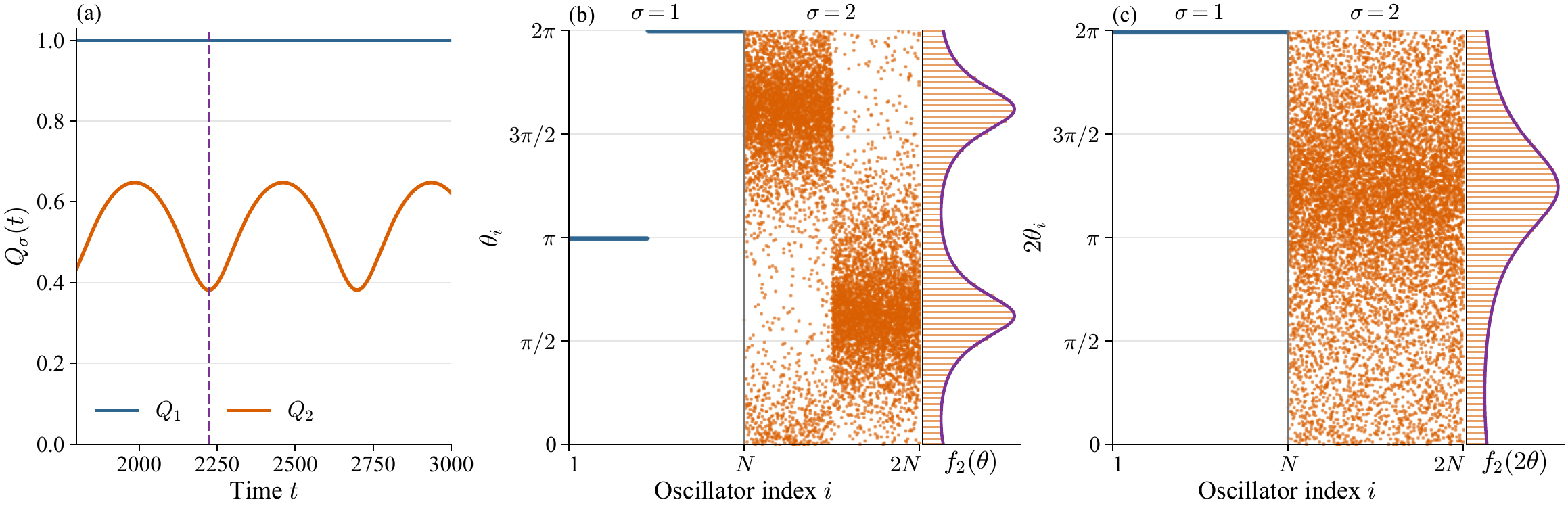}
\caption{Microscopic view of a breathing chimera state, obtained by numerical integration of Eq.~\eqref{eq:eq_of_motion} with $N_1=N_2=10\,000$, $Q_1(0)=1$, $Q_2(0)=0.60$, $\eta_1=0.55$, $\eta_2=0.50$, $\alpha=\pi/4$, and $\mu=0.50$.
(a) Time evolution of $Q_1$ and $Q_2$. The purple dashed line indicates the selected minimum of $Q_2$ at $t=2223.5$.
(b) Oscillator phases $\theta_i$ at the selected time, showing a phase-locked bimodal distribution in Group~1, while the phases in Group~2 are spread over a broad range.
(c) The same snapshot represented by $2\theta_i\bmod 2\pi$. The two peaks of Group~1 separated by $\pi$ are mapped onto the same phase, whereas the phases in Group~2 remain broadly distributed. The smooth curves show the phase density $f_2(\theta)$ obtained from the OA ansatz in Eq.~(S7) in (b) and the corresponding wrapped doubled-phase density $f_2(2\theta)$ in (c); both agree with the observed histograms. The corresponding temporal evolution is shown in Supplementary Video~S2.}
\label{fig:phase_distribution}
\end{figure*}
\FloatBarrier


{\it Reduced dynamics ---}
By substituting the modified OA ansatz into the continuity equation for the phase density $f$ and collecting the resulting Fourier harmonics, we reduce the infinite-dimensional dynamics to a closed set of equations for the macroscopic variables (see Supplemental Material Sec.~{S1}.D for details). Using $a_\sigma = r_\sigma e^{-{\rm i}\psi_\sigma}$ in Eq.~\eqref{eq:f_OAform}, the system is fully described by the amplitude $r_\sigma$ and the phase $\psi_\sigma$ of each group. Focusing on the interaction between the two groups, it is convenient to introduce the phase difference $\phi \equiv \psi_1 - \psi_2$. Under the asymmetric initialization described above, with $r_1=1$, $\eta_1\neq0.5$, and $\eta_2=0.5$, the dynamics simplifies to a two-dimensional system governing the evolution of $r\equiv r_2$ and $\phi$ (with the shorthand $\eta \equiv \eta_1$):
\begin{align}
\dot{r}
&=
\frac{1}{2r}(1 - r^4)\,\mu^2 (2\eta - 1)^2 \cos(2\phi + \alpha)~,  \label{eq:rdot}\\
\dot{\phi}
&=
-\frac{1}{2r^2}(1 + r^4)\,\mu^2 (2\eta - 1)^2 \sin(2\phi + \alpha)+ (2\eta - 1)^2 \sin\alpha~. \label{eq:phidot}
\end{align}
For given initial conditions and parameters $\mu$, $\eta$, and $\alpha$, it provides a minimal description of the interplay between coherence ($r$) and phase difference ($\phi$), allowing direct analysis of steady states, stability, and oscillatory behavior.

\begin{figure*}[!htbp]
\centering
\includegraphics[width=\textwidth]{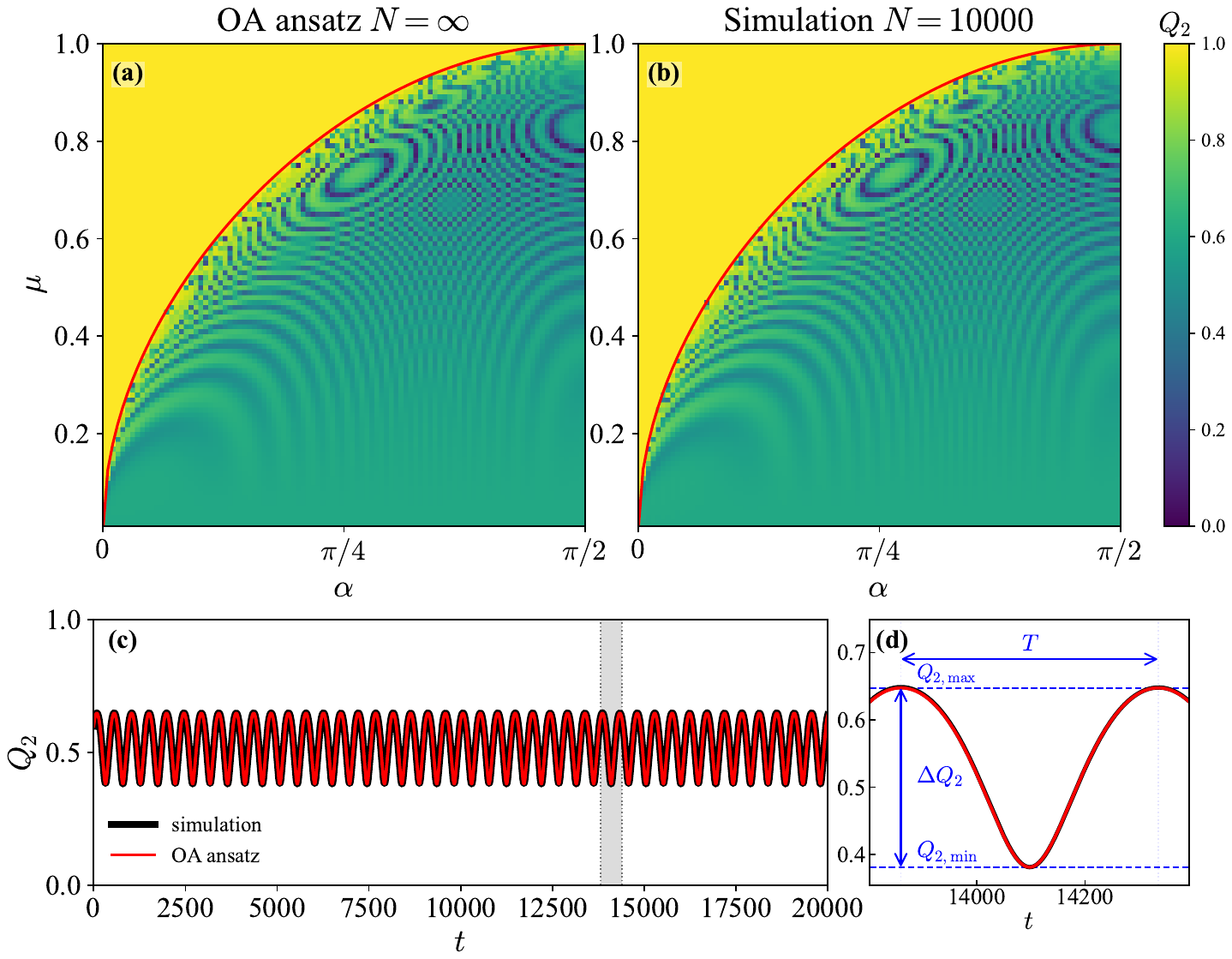}
\caption{
Phase diagrams in the $(\alpha,\mu)$ space and temporal dynamics of $Q_2$.
For all panels, the initial conditions are $Q_1(0)=1$, $Q_2(0)=0.60$, $\eta_1=0.55$, and $\eta_2=0.50$. (a) Instantaneous value of $Q_2$ at $t=20000$ in the thermodynamic limit 
($N_\sigma \rightarrow \infty$). (b) Corresponding result from numerical simulations with $N_1=N_2=10000$. Color represents the instantaneous value of $Q_2$ after the transient, which varies in time due to the breathing dynamics. The theoretical boundary (red solid line), given by $\mu=\sqrt{\sin\alpha}$, is in agreement with the boundary of the breathing chimera state region obtained from numerical simulations. See Supplementary Video S1 for the temporal evolution of the phase diagrams in panels (a,b). (c) Time evolution of $Q_2$ for $(\alpha,\mu)=(\pi/4,0.5)$. The numerical simulation result (thick black line) agrees well with the theoretically predicted thermodynamic-limit result (thin red line).
(d) Zoomed-in view of the shaded time interval in panel (c), showing one oscillation period.
The theoretical amplitude and period, $\Delta Q_2\simeq0.266$ and $T\simeq475.0$, agree with the corresponding values measured from the simulation, $\Delta Q_2\simeq0.266$ and $T\simeq474.9$.
The absolute differences between theory and simulation are approximately $5.1\times10^{-4}$ for $\Delta Q_2$ and $6.4\times10^{-2}$ for $T$.
The finite-size scaling of the residual discrepancy between the numerical simulation and the OA prediction is presented in Supplemental Material Sec.~S5.}

\label{fig:fig2}
\end{figure*}

{\it Fixed points ---}
The reduced system admits steady-state solutions determined by $\dot{r}=0$ and $\dot{\phi}=0$. Fixed points arise either at $r=1$ (fully synchronized state) or when $\cos(2\phi+\alpha)=0$ from  $\dot{r}=0$. The latter corresponds to partially synchronized states, which we identify as chimera states as shown in Fig.~\ref{fig:fig2}. Together with $\dot{\phi}=0$, the condition for the synchronized regime in the second group $\sigma=2$ is given by $\mu \ge \sqrt{\sin\alpha}$ (for full synchronization) and $\mu < \sqrt{\sin\alpha}$ (for partial synchronization),
which defines a critical threshold for the emergence of chimera states when equality holds: above this threshold the system converges to full synchronization, resulting in $Q_1=1$ and $Q_2=1$, whereas below it chimera states ($Q_1=1$ and $Q_2<1$ equivalent to $r<1$) become accessible, as portrayed in Figs.~\ref{fig:fig2}(a)~and~\ref{fig:fig2}(b). 
The nonlinear dynamics in this regime is analyzed exactly in the following section using a Riccati-type formulation and its conserved invariant.
The simulation results for $Q_2$ in the regime $\mu < \sqrt{\sin \alpha}$ agree with the prediction obtained from the value of $r$ given by Eqs.~\eqref{eq:order}, \eqref{eq:rdot}, and~\eqref{eq:phidot} in Fig.~\ref{fig:fig2}(c), showing the corresponding {closed-orbit behavior}.

\FloatBarrier

\section{Dynamics in Complex Plane and Geometric Interpretation}
The breathing chimera appears as persistent oscillations in the reduced variables. To interpret this motion geometrically, we combine the reduced variables into a single complex order parameter $Z = r^2 e^{{\rm i}\Phi}$ with $\Phi = 2\phi + \alpha$. The dynamics then reduces to the Riccati-type equation~\cite{Marvel2009}
\begin{equation}
\dot{Z} = A\left(1-Z^2\right)+2{\rm i}BZ~,
\end{equation}
where $A=\mu^2(1-2\eta)^2$ and $B=\sin\alpha\,(1-2\eta)^2$.

\begin{figure*}[!hb]
\centering
\includegraphics[width=\textwidth]{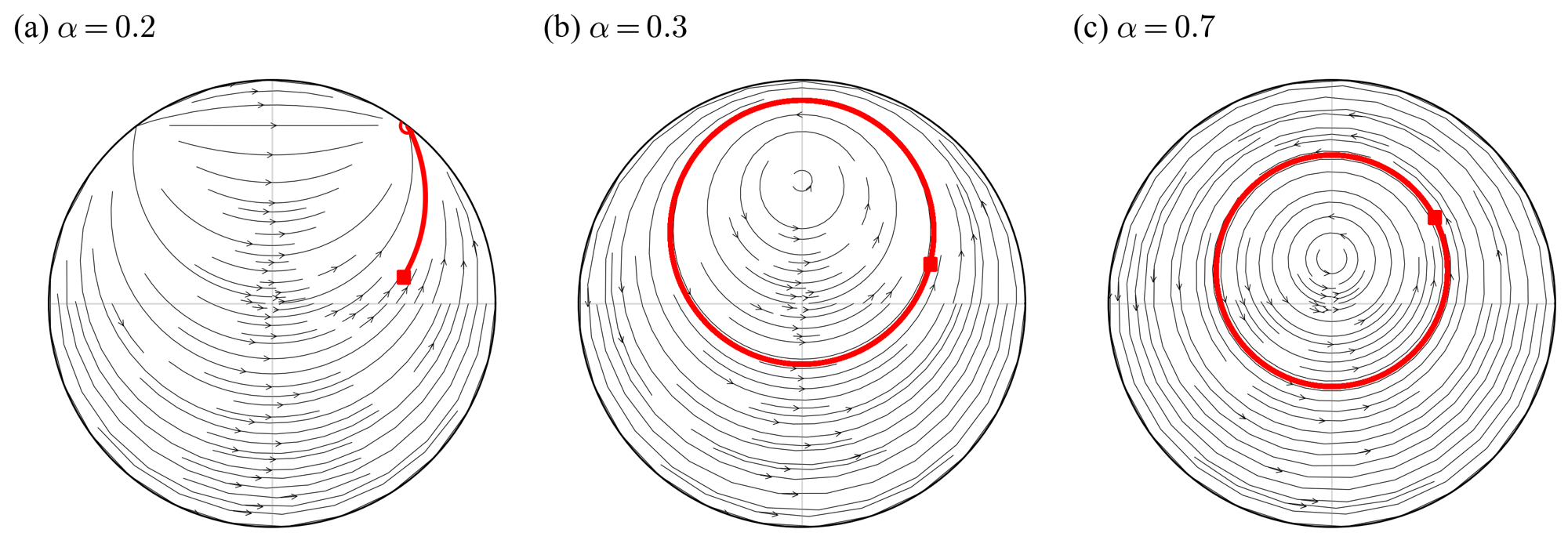}
\caption{Phase portraits in the complex plane, shown in polar coordinates, for $\mu=0.5$.
The trajectories (red curves), initialized at $Q_2(0)=0.6$, either approach the fully synchronized state ($Q_2=1$) or evolve along neutrally stable closed orbits depending on $\alpha$.
For all panels, the remaining initial conditions are $Q_1(0)=1$, $\eta_1=0.55$, and $\eta_2=0.50$.
(a) For $\alpha=0.2$, the trajectory approaches the fully synchronized state ($Q_2=1$).
(b,c) For $\alpha=0.3$ and $\alpha=0.7$, the trajectories evolve along neutrally stable closed orbits, corresponding to breathing chimera states.}
\label{fig:fig3}
\end{figure*}

This compact form reveals that the breathing chimera is governed by an exactly tractable low-dimensional structure. The Riccati equation admits a conserved invariant. 
In the breathing regime, its physical level sets form closed periodic orbits in the complex plane. Writing $Z=X+{\rm i}Y$, the invariant takes the form
\begin{equation}
X^2+\left(Y-\frac{A}{2C_0}\right)^2
=
\frac{A^2-4C_0(B-C_0)}{4C_0^2}~,
\end{equation}
where $C_0$ is the conserved value of the invariant determined by the initial condition of the complex variable $Z$ for given system parameters $A$ and $B$ (see Supplemental Material Sec.~{S3}.B for details). This representation exposes the invariant structure underlying the breathing motion and allows the reduced trajectory to be interpreted as a circular orbit in the complex plane. Figure~\ref{fig:fig3} illustrates how the reduced dynamics converges to the fully synchronized state or {evolves along neutrally stable closed orbits} depending on $\alpha$.

The oscillatory dynamics of the breathing chimera can be characterized by its period and amplitude. The exact nonlinear solution of the Riccati-type equation {shows that every non-equilibrium closed orbit in the physical domain has the period}
$T/\pi=(1-2\eta)^{-2}(\sin^2\alpha-\mu^4)^{-1/2}$. 
Different initial conditions select closed orbits with different
amplitudes but do not change their period. Thus, the family of breathing
orbits is isochronous. This expression shows that the period depends only on the system parameters and diverges as $\mu\to\sqrt{\sin\alpha}$, indicating critical slowing down near the bifurcation point.
In contrast, the oscillation amplitude depends on the initial condition through the invariant of motion. For example, for symmetric initial states, the amplitude scales as
$\Delta Q \sim {\mu^2}/{\sin\alpha}$ [Fig.~\ref{fig:fig2}(d)],
demonstrating that the extent of desynchronization is controlled by both coupling strength and phase lag. See Supplemental Material Sec.~{S4} for detailed derivations.

\FloatBarrier

\section{Discussion and Conclusions}
In this work, we demonstrated that chimera states can emerge in an oscillator system governed solely by triadic interactions, without any pairwise coupling. A central feature of the dynamics is the intrinsic $\pi$-symmetry generated by the triadic coupling, which supports bimodal phase distributions. To capture this structure, we developed a modified Ott–Antonsen ansatz incorporating an asymmetry parameter that characterizes the imbalance between two anti-phase subpopulations, enabling an exact dimensional reduction of the microscopic dynamics into a low-dimensional macroscopic system.

The exact reduction further allows us to derive the conditions for the emergence of chimera states and 
identify the bifurcation threshold separating synchronized and oscillatory regimes. Reformulating the reduced dynamics in terms of a complex order parameter further revealed that the breathing chimera corresponds to persistent periodic orbits with a simple geometric structure in the complex plane. The role of triadic coupling is thus not simply to replace pairwise coupling, but to impose an intrinsic $\pi$-symmetry that supports bimodal collective states: within this bimodal structure, an initial imbalance between the two anti-phase subpopulations allows one group to remain fully synchronized, while the other exhibits partial synchronization with persistent oscillations. We further examined the dependence on initial conditions. The breathing chimera state is observed both from the low-coherence initial condition $Q_1(0)=0.1$ and $Q_2(0)=0$ and under independent perturbations of the initial phases. In the perturbed cases, Group~1 recovers $Q_1\simeq1$, while Group~2 retains persistent oscillations in $Q_2(t)$ over the observation windows considered (see Supplemental Material Sec.~S6 and Supplementary Video~S3).

These results establish an analytically tractable mechanism for chimera formation driven by purely higher-order interactions, suggesting more broadly that many-body coupling can support symmetry-broken collective dynamics distinct from that generated by conventional pairwise interactions, with potential implications for biological and neural systems where higher-order interactions may play an intrinsic role~\cite{Parastesh2021,Boccaletti2023}.
The present analysis focuses on identical oscillators with all-to-all purely triadic interactions and prescribed asymmetric initial conditions; these assumptions make the model analytically tractable but leave open how the mechanism is affected by frequency heterogeneity, noise, sparse hypergraph topology, and mixed pairwise–triadic coupling. Addressing these extensions would clarify the robustness of breathing chimera states driven by purely higher-order interactions in more realistic networked systems.

\section{Acknowledgments} \label{sec:acknowledgements}
This work was supported by the National Research Foundation (NRF) of Korea through Grant Nos. RS-2023-00279802 (S.Y., B.K.), RS-2024-00341317 (M.J.L.), and RS-2026-25488703 (S.-W.S.). This work was also partly supported by Korea Research Institute for defense Technology  planning and advancement  (KRIT) - Grant funded  by  Defense  Acquisition  Program Administration  (DAPA),  South  Korea  (KRIT-CT-23-026, Integrated  Underwater  Surveillance  Research  Center  for Adapting Future Technologies, 2023–2029). We thank APCTP, Pohang, Korea, for their hospitality during the Topical Research Program [APCTP-2025-T04], from which this work greatly benefited.

\printcredits

\bibliographystyle{elsarticle-num}
\bibliography{refs}

\end{document}


\title{Supplemental Material for ``Breathing chimera states from purely triadic interactions''}
\author{Sudo Yi}
\affiliation{CCSS, KI for Grid Modernization, Korea Institute of Energy Technology, Korea}
\author{Gug Young Kim}
\affiliation{Department of Applied Physics, Hanyang University ERICA, Korea}
\author{Mi Jin Lee}
\email{mijinlee@pusan.ac.kr}
\affiliation{Department of Physics, Pusan National University, Korea}
\author{S.-W. Son}
\email{sonswoo@hanyang.ac.kr}
\affiliation{Department of Applied Physics, Hanyang University ERICA, Korea}
\author{B.~Kahng}
\email{bkahng@kentech.ac.kr}
\affiliation{CCSS, KI for Grid Modernization, Korea Institute of Energy Technology, Korea}
\date{\today}

\maketitle
\tableofcontents

\section{Bimodal Ott--Antonsen reduction}
\subsection{Mean-field description of the triadic interaction model}
We consider two groups of identical oscillators interacting solely through triadic coupling. In the thermodynamic limit, the phase distribution of group $\sigma$ evolves under the conservation of oscillators in phase space, so that it is described by a probability density $f_\sigma(\theta,t)$, which obeys the continuity equation
\begin{equation}
\frac{\partial f_\sigma}{\partial t} + \frac{\partial}{\partial\theta} \left(f_\sigma v^\sigma\right)=0.
\label{eq:continuity_eq}
\end{equation}
According to Eq.{~(1)} in the main text, the velocity is given by
\begin{equation}
v^\sigma(\theta,t)=\omega+\frac{1}{2{\rm i}} \left(H_\sigma e^{-{\rm i}(2\theta+\alpha)}-H_\sigma^* e^{{\rm i}(2\theta+\alpha)}\right),
\end{equation}
where the mean field $H_\sigma$ is defined as
\begin{equation}
H_\sigma=\sum_{\sigma',\sigma''}
K_{\sigma\sigma'\sigma''}
Z_{1,\sigma'}Z_{1,\sigma''}.
\label{eq:sm_MF}
\end{equation}
The complex order parameters are defined as the conventional Kuramoto order parameter $Z_{1, \sigma}=R_{\sigma} e^{{\rm i}\psi_{\sigma}}$ and the Daido order parameter  $Z_{2, \sigma}=Q_{\sigma} e^{{\rm i}\chi_{\sigma}}$, and $Z_{1, \sigma}$ appears only in Eq.~\eqref{eq:sm_MF}.

\subsection{Modified bimodal Ott–Antonsen ansatz}
\label{subsec:OA}
Because the triadic interaction contains the term $-2\theta_i^\sigma$ in Eq.{~(1) of the main text}, the phase dynamics possesses an intrinsic $\pi$-symmetry, which permits for bimodal phase distributions. To capture this structure, we expand the phase density as
\begin{equation}
f_\sigma(\theta,t)=\frac{1}{2\pi} \left[ 1+\sum_{n=1}^\infty \left( \hat f_\sigma^{(n)}e^{{\rm i}n\theta}+\text{c.c.} \right) \right].
\end{equation} 
We then decompose the distribution into two subpopulations,
$f_\sigma=\eta_\sigma f_{\sigma,a}+(1-\eta_\sigma)f_{\sigma,b}$,
with
\begin{equation}
f_{\sigma,a} = \frac{1}{2\pi} \left[ 1+\sum_{n=1}^{\infty} \left( \hat{f}_{\sigma,a}^{(n)}e^{{\rm i}n\theta}+\text{c.c.} \right) \right],
\qquad
f_{\sigma,b} = \frac{1}{2\pi} \left[ 1+\sum_{n=1}^{\infty} \left( \hat {f}_{\sigma,b}^{(n)}e^{{\rm i}n\theta}+\text{c.c.} \right) \right].
\end{equation}
We assume that the two peaks are symmetrically located in phase space, i.e., $f_{\sigma,a}(\theta,t)=f_{\sigma,b}(\theta-\pi,t)$, reflecting the intrinsic $\pi$-shift symmetry.

Applying the Ott--Antonsen (OA) ansatz separately to the two subpopulations, we set $\hat{f}_{\sigma,a}^{(n)}= a_\sigma^n$ and $\hat{f}_{\sigma,b}^{(n)}=b_\sigma^{n}$, implying $b_\sigma=-a_\sigma$ due to the $\pi$-shift symmetry, which gives
\begin{equation}
f_\sigma = \frac{1}{2\pi} \left( 1+\left[\sum_{n=1}^{\infty}a_\sigma^{2n}e^{{\rm i}2n\theta} + (2\eta_\sigma-1)\sum_{n=1}^{\infty}a_\sigma^{2n-1}e^{{\rm i}(2n-1)\theta} \right] +\text{c.c.}\right).
\end{equation}
{At $t=0$, this expression specifies the initial phase distribution once $a_\sigma(0)=r_\sigma(0)e^{-{\rm i}\psi_\sigma(0)}$ and $\eta_\sigma$ are prescribed; the particular choices used here are given in Sec.~S1.C.}

Summing the geometric series yields the closed-form density
\begin{equation}
f_\sigma =
\frac{1}{2\pi}
\left[
1 +
\frac{
2\left[
(2\eta_\sigma-1)r_\sigma(1-r_\sigma^2)\cos\left(\theta-\psi_\sigma\right)
+r_\sigma^2\cos2\left(\theta-\psi_\sigma\right)-r_\sigma^4
\right]
}{
1-2r_\sigma^2\cos2\left(\theta-\psi_\sigma\right)+r_\sigma^4
}
\right].
\end{equation}
Writing $a_\sigma=r_\sigma e^{-{\rm i}\psi_\sigma}$,
the first- and second-harmonic moments become
\begin{equation}
Z_{1,\sigma}\left(t\right)
=
\int_{-\pi}^{\pi}
f_\sigma(\theta,t)e^{{\rm i}\theta}d\theta
=
(2\eta_\sigma-1)a_\sigma^*,\qquad
Z_{2,\sigma}\left(t\right)
=
\int_{-\pi}^{\pi}
f_\sigma(\theta,t)e^{2{\rm i}\theta}d\theta
=
(a_\sigma^*)^2.
\end{equation}
Therefore, the magnitudes of Kuramoto and Daido order parameters are obtained as
\begin{equation}
R_\sigma=\left|2\eta_\sigma-1\right|r_\sigma,
\qquad
Q_\sigma=r_\sigma^2,
\label{eq:R_Q}
\end{equation}
{where the corresponding order parameters are defined as 
\begin{equation}
   R_\sigma e^{{\rm i}\psi_\sigma}=
\frac{1}{N_\sigma}\sum_{j=1}^{N_\sigma} e^{{\rm i}\theta_j^\sigma}, \qquad Q_\sigma e^{{\rm i}\chi_\sigma} =
\frac{1}{N_\sigma}\sum_{j=1}^{N_\sigma} e^{{\rm i}2\theta_j^\sigma}
\end{equation}}
respectively.

The parameter $\eta_\sigma$ is fixed by the initial imbalance between the two anti-phase subpopulations. 
Within the present bimodal OA construction, the two subpopulations evolve with the same macroscopic variable $a_\sigma$ up to the $\pi$ shift, so that their relative weights are conserved. 
Thus, $\eta_\sigma$ is not an additional dynamical variable but a parameter determined by the initial phase distribution.

\subsection{Initial Conditions}
Higher-order interaction systems are strongly multistable, so their macroscopic dynamics depends sensitively on the initial condition. Since
$R_\sigma=\left|2\eta_\sigma-1\right|r_\sigma$,
the parameter $\eta_\sigma$ directly controls the initial imbalance between the two anti-phase subpopulations. In particular, $\eta_\sigma=0.5$ corresponds to a symmetric bimodal state with $R_\sigma=0$.

\begin{figure}
    \centering
    \includegraphics[width=0.8\linewidth]{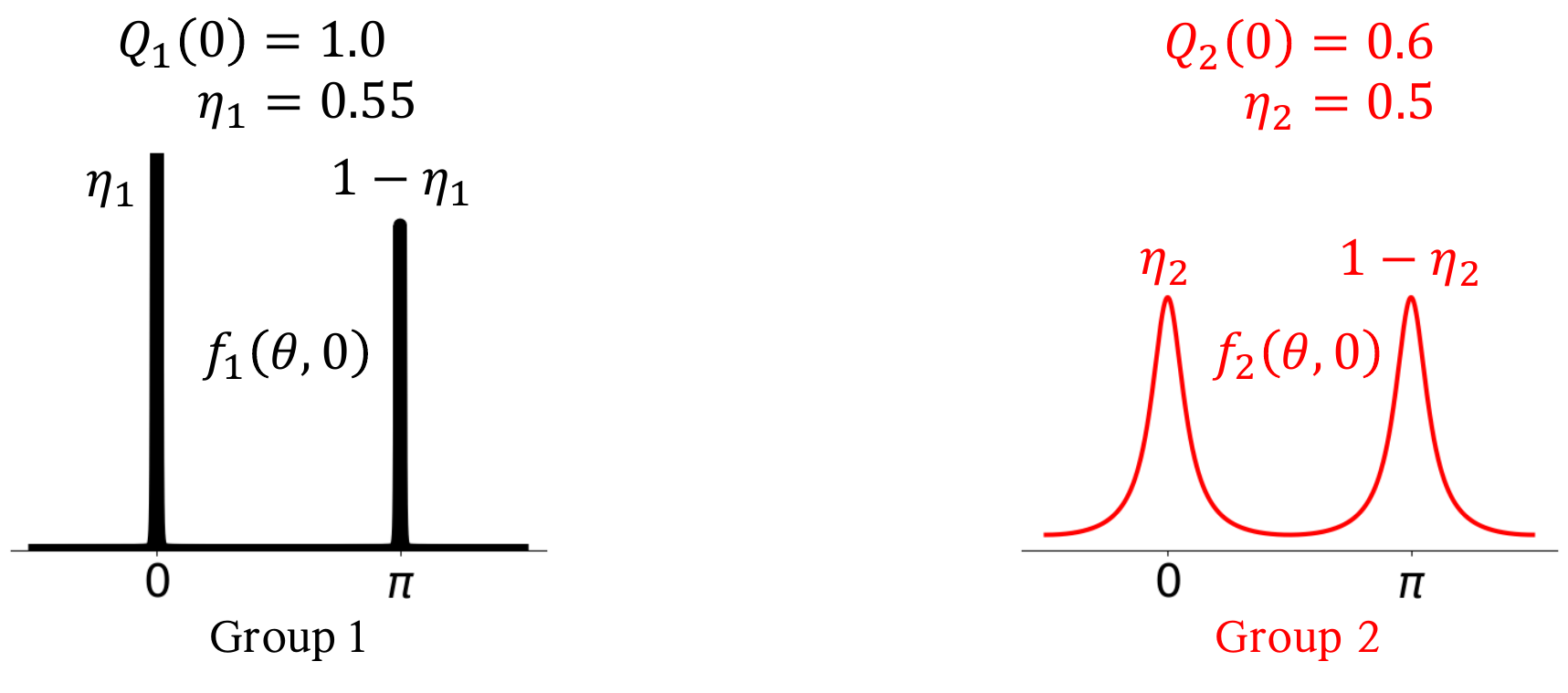}
    \caption{Initial phase distributions of the two groups modeling the unihemispheric sleep scenario. Group 1 (sleeping hemisphere) exhibits a prescribed asymmetry between the two $\pi$-shifted subpopulations, forming an asymmetric synchronized bimodal state that is a fully synchronized ($Q_1 = 1$, $\eta_1 = 0.55$).} Conversely, Group 2 (awake hemisphere) preserves the inherent $\pi$-symmetry of triadic interactions, maintaining a symmetric partially synchronized bimodal state ($Q_2 < 1$, $\eta_2 = 0.5$). This asymmetric initialization serves as the initial condition for the emergence of chimera states. 
    \label{fig:fig_initial_condition}
\end{figure}

To induce chimera states, we consider an asymmetric initialization inspired by unihemispheric sleep, as illustrated in Fig.~\ref{fig:fig_initial_condition}. Group 1 is prepared in an asymmetric synchronized state, $a_1=1$ and $\eta_1\neq0.5$, so that
\begin{equation}
Q_1=1,\qquad R_1\neq 0.
\end{equation}
The other group remains symmetric and partially synchronized, $a_2=\sqrt{0.6}$ and $\eta_2=0.5$, giving
\begin{equation}
Q_2=0.6,\qquad R_2=0.
\end{equation} 

From Eq.~(\ref{eq:R_Q}), this choice corresponds to $r_1=1$, while $\eta_1\neq0.5$ and $\eta_2=0.5$ specify the relative weights of the two anti-phase subpopulations in Groups 1 and 2, respectively.


For the finite-$N$ simulations, the initial phases were generated by
deterministic regular inverse-CDF sampling. For Group~1, $r_1=1$
reduces the two component densities to point masses, and $\eta_1N_1$
and $(1-\eta_1)N_1$ oscillators were placed at $\theta=0$ and $\pi$,
respectively. For Group~2, we define
\begin{equation}
F_{2,a}(\theta)=\int_0^\theta f_{2,a}(\vartheta,0)\,d\vartheta .
\end{equation}
The first $N_2/2$ phases were chosen as $\theta_m^{\sigma=2}=F_{2,a}^{-1}\left(\frac{m-1/2}{N_2/2}\right)$, $m=1,\ldots,N_2/2$, and the remaining phases were assigned as $\theta_{m+N_2/2}^{\sigma=2} =\theta_m^{\sigma=2}+\pi$.

\subsection{Derivation of the reduced macroscopic dynamics}
Substituting the modified OA ansatz with $a_\sigma=r_\sigma e^{-{\rm i}\psi_\sigma}$ into the continuity equation Eq.~(\ref{eq:continuity_eq}) gives
\begin{equation}
\dot{a}_\sigma =-{\rm i}\omega r_\sigma e^{-{\rm i}\psi_\sigma} -\frac{1}{2}\left( H_\sigma r_\sigma^3 e^{-3i\psi_\sigma}e^{-{\rm i}\alpha} -H_\sigma^*\frac{1}{r_\sigma}e^{{\rm i}\psi_\sigma}e^{{\rm i}\alpha} \right).    
\end{equation}

After substituting the order parameters and separating real and imaginary parts, we obtain
\begin{align}
\dot{r}_\sigma
=&\;
\frac{1}{2r_\sigma}(1-r_\sigma^4)\Big[ (2\eta_\sigma-1)^2 r_\sigma^2 \cos\alpha + 2\mu(2\eta_\sigma-1)(2\eta_{\sigma'}-1) r_\sigma r_{\sigma'} \cos(\psi_\sigma-\psi_{\sigma'}+\alpha) \notag\\
&+ \mu^2(2\eta_{\sigma'}-1)^2 r_{\sigma'}^2 \cos\!\big(2(\psi_\sigma-\psi_{\sigma'})+\alpha\big)
\Big],
\\[6pt]
\dot{\psi}_\sigma
=&\; \omega
-\frac{1}{2r_\sigma^2}(1+r_\sigma^4)
\Big[
(2\eta_\sigma-1)^2 r_\sigma^2 \sin\alpha + 2\mu(2\eta_\sigma-1)(2\eta_{\sigma'}-1) r_\sigma r_{\sigma'}
\sin(\psi_\sigma-\psi_{\sigma'}+\alpha) \notag\\
&+ \mu^2(2\eta_{\sigma'}-1)^2 r_{\sigma'}^2
\sin\!\big(2(\psi_\sigma-\psi_{\sigma'})+\alpha\big)
\Big].
\end{align}

Defining the phase difference $\phi=\psi_\sigma-\psi_{\sigma'}$, which suffices due to invariance under a global phase shift, and considering the asymmetric initialization $\eta_2=0.5$, $r_1=1$, we reduce the dynamics to
\begin{align}
\dot r &=
\frac{1}{2r}(1-r^4)\mu^2(2\eta-1)^2\cos(2\phi+\alpha),\\
\dot\phi&=
-\frac{1}{2r^2}(1+r^4)\mu^2(2\eta-1)^2\sin(2\phi+\alpha)
+(2\eta-1)^2\sin\alpha,
\end{align}
where $r\equiv r_2$ and $\eta\equiv \eta_1$ for notational convenience.

\section{Fixed points and stability analysis}
\subsection{Existence of fixed points}
Fixed points satisfy $\dot r=0$ and $\dot\phi=0$. From the amplitude equation, either
$r=1$
or
$\cos(2\phi+\alpha)=0$.

For $\cos(2\phi+\alpha)=0$, one has $\sin(2\phi+\alpha)=\pm1$, and substitution into the phase equation yields
\begin{equation}
r^4 \mp \frac{2r^2\sin\alpha}{\mu^2}
+\left(\frac{\sin\alpha}{\mu^2}\right)^2
=
\left(\frac{\sin\alpha}{\mu^2}\right)^2-1.
\end{equation}
Hence,
\begin{equation}
r^2
=
\pm\frac{\sin\alpha}{\mu^2}
\pm
\sqrt{
\left(\frac{\sin\alpha}{\mu^2}\right)^2-1
},
\end{equation}
which requires $\mu\le \sqrt{\sin\alpha}$.

For $r=1$, the phase equation gives
\begin{equation}
    \sin(2\phi+\alpha)=\frac{\sin\alpha}{\mu^2},
\end{equation}
which requires
$\mu\ge \sqrt{\sin\alpha}$.

\subsection{Linear stability}
The Jacobian is
\begin{equation}
J=
\begin{pmatrix}
\partial\dot r/\partial r & \partial\dot r/\partial\phi\\
\partial\dot \phi/\partial r & \partial\dot\phi/\partial\phi
\end{pmatrix}
=
\frac{\mu^2(2\eta-1)^2}{2r^3}
\begin{pmatrix}
-r(1+3r^4)\cos(2\phi+\alpha)
&
-2r^2(1-r^4)\sin(2\phi+\alpha)
\\[1.5ex]
2(1-r^4)\sin(2\phi+\alpha)
&
-2r(1+r^4)\cos(2\phi+\alpha)
\end{pmatrix}.
\end{equation}

For $r\neq1$ giving $\cos(2\phi+\alpha)=0$, the trace vanishes and the determinant is positive,
\begin{equation}
\mathrm{Tr}(J)=0,\qquad
\mathrm{Det}(J)=4(2\eta-1)^4(\sin^2\alpha-\mu^4)>0.
\end{equation}
Thus, the eigenvalues are purely imaginary. The nonlinear dynamics is analyzed separately using the exact Riccati-type equation in Sec.~S3. For $r=1$, the Jacobian becomes diagonal and the fixed point is a degenerate sink or source depending on the sign of $\cos(2\phi+\alpha)$.

\section{Dynamics in the complex plane}

\subsection{Reduction to a Riccati equation}
We now rewrite the reduced dynamics in terms of a single complex variable. 
For notational simplicity, we define
\begin{equation}
A=\mu^2(2\eta-1)^2,
\qquad
B=\sin\alpha(2\eta-1)^2 .
\end{equation}
Then the reduced equations become
\begin{align}
\dot r &= \frac{A}{2r}(1-r^4)\cos(2\phi+\alpha), \label{eq:r_simple}\\
\dot\phi &= -\frac{A}{2r^2}(1+r^4)\sin(2\phi+\alpha)+B. \label{eq:phi_simple}
\end{align}

Introducing
\begin{equation}
\Phi=2\phi+\alpha,
\qquad
Z=r^2 e^{{\rm i}\Phi},
\end{equation}
we obtain
\begin{align}
\dot Z
&=
\frac{d}{dt}\left(r^2e^{{\rm i}\Phi}\right)  \notag\\
&=
\left(2r\dot r+2{\rm i}r^2\dot\phi\right)e^{{\rm i}\Phi}.
\end{align}
Substituting Eqs.~\eqref{eq:r_simple} and \eqref{eq:phi_simple} gives
\begin{align}
\dot Z
&=
\left[
A(1-r^4)\cos\Phi
-{\rm i}A(1+r^4)\sin\Phi
+2{\rm i}Br^2
\right]e^{{\rm i}\Phi} \notag\\
&=
\left[
Ae^{-{\rm i}\Phi}
-Ar^4e^{{\rm i}\Phi}
+2{\rm i}Br^2
\right]e^{{\rm i}\Phi}.
\end{align}
Using $Z=r^2e^{{\rm i}\Phi}$, this reduces to
\begin{equation}
\dot Z=A(1-Z^2)+2{\rm i}BZ.
\label{eq:riccati}
\end{equation}
Thus, the reduced macroscopic dynamics is represented by a Riccati-type equation.

\subsection{Invariant manifold and circular orbit}
To obtain the geometric structure of the orbit, we write
\begin{equation}
Z=X+{\rm i}Y.
\end{equation}
Separating Eq.~\eqref{eq:riccati} into real and imaginary parts gives
\begin{align}
\dot X &= A(1-X^2+Y^2)-2BY, \label{eq:Xdot}\\
\dot Y &= 2X(B-AY). \label{eq:Ydot}
\end{align}
Taking the ratio $dY/dX=\dot Y/\dot X$, we obtain
\begin{equation}
2X(B-AY)dX-\left[A(1-X^2+Y^2)-2BY\right]dY=0.
\end{equation}
Multiplying by the integrating factor $(1-X^2-Y^2)^{-2}$, we obtain an exact equation, whose invariant is
\begin{equation}
H(X,Y)=\frac{B-AY}{1-X^2-Y^2}=C_0,
\label{eq:invariant}
\end{equation}
where $C_0$ is determined by the initial condition.

Rearranging Eq.~\eqref{eq:invariant}, we obtain
\begin{equation}
X^2+Y^2-\frac{A}{C_0}Y
=
1-\frac{B}{C_0},
\end{equation}
or equivalently,
\begin{equation}
X^2+\left(Y-\frac{A}{2C_0}\right)^2
=
\frac{A^2-4C_0(B-C_0)}{4C_0^2}.
\label{eq:circle_orbit}
\end{equation}
Therefore, the trajectory of $Z$ is a circle in the complex plane, with center
\begin{equation}
\left(0,\frac{A}{2C_0}\right)
\end{equation}
and radius
\begin{equation}
\frac{\sqrt{A^2-4C_0(B-C_0)}}{2|C_0|}.
\end{equation}
This circular orbit corresponds to the breathing chimera state in the reduced macroscopic dynamics.
Since $C_0$ is conserved, these circular trajectories form a neutrally stable family of closed periodic orbits.

\section{Period and amplitude of breathing chimera states}

The period of the breathing chimera can be obtained from the exact nonlinear solution of the Riccati-type equation. For $B^2>A^2$, we define
\begin{equation}
\Omega_R=\sqrt{B^2-A^2},
\qquad
Z_\pm=\frac{{\rm i}(B\pm\Omega_R)}{A}.
\end{equation}
The exact solution of Eq.~\eqref{eq:riccati} satisfies
\begin{equation}
\frac{Z(t)-Z_+}{Z(t)-Z_-}
=
C e^{-2{\rm i}\Omega_Rt},
\end{equation}
where $C$ is determined by the initial condition. Although $C$ selects the oscillation amplitude, it does not affect the frequency $\Omega_R$. Therefore, the exact period is
\begin{equation}
T=\frac{\pi}{\Omega_R}
=
\frac{\pi}
{(2\eta-1)^2\sqrt{\sin^2\alpha-\mu^4}}.
\label{eq:period}
\end{equation}
Thus, the period is independent of the oscillation amplitude.

The oscillation amplitude is determined by the invariant $C_0$. For mathematical convenience and consistency with the scaled parameters, we define a scaled invariant $\tilde{C}_0 \equiv C_0 / (2\eta-1)^2$. Since
\begin{equation}
Z=r^2e^{{\rm i}\Phi}=Qe^{{\rm i}\Phi},
\end{equation}
the radial distance of the orbit in the complex plane corresponds to the Daido order parameter $Q$.
From the circular orbit in Eq.~\eqref{eq:circle_orbit}, by substituting $A$ and $B$, the maximum and minimum values of $Q$ can be expressed compactly in terms of $\tilde{C}_0$ as
\begin{align}
Q_{\max}
&=
\frac{\mu^2+\sqrt{\mu^4-4\tilde{C}_0(\sin\alpha-\tilde{C}_0)}}{2\tilde{C}_0},
\\
Q_{\min}
&=
\left|
\frac{-\mu^2+\sqrt{\mu^4-4\tilde{C}_0(\sin\alpha-\tilde{C}_0)}}{2\tilde{C}_0}
\right|.
\end{align}
Thus, the amplitude of the breathing chimera is
\begin{equation}
\Delta Q=Q_{\max}-Q_{\min}.
\end{equation}

For the special case $Q(0)=0$ (where $X=0$ and $Y=0$), the definition of the invariant yields $C_0 = B = \sin\alpha(2\eta-1)^2$, which implies that the scaled invariant simply becomes $\tilde{C}_0=\sin\alpha$.
Then, the extrema simplify to
\begin{equation}
Q_{\max}=\frac{\mu^2}{\sin\alpha},
\qquad
Q_{\min}=0,
\end{equation}
and hence the amplitude is given by
\begin{equation}
\Delta Q=\frac{\mu^2}{\sin\alpha}.
\label{eq:amplitude}
\end{equation}
Unlike the period, which depends only on the system parameters, the amplitude depends on the initial condition through $\tilde{C}_0$.

\section{Numerical validation and finite-size scaling}
\begin{figure}
    \centering
    \includegraphics[width=0.45\linewidth]{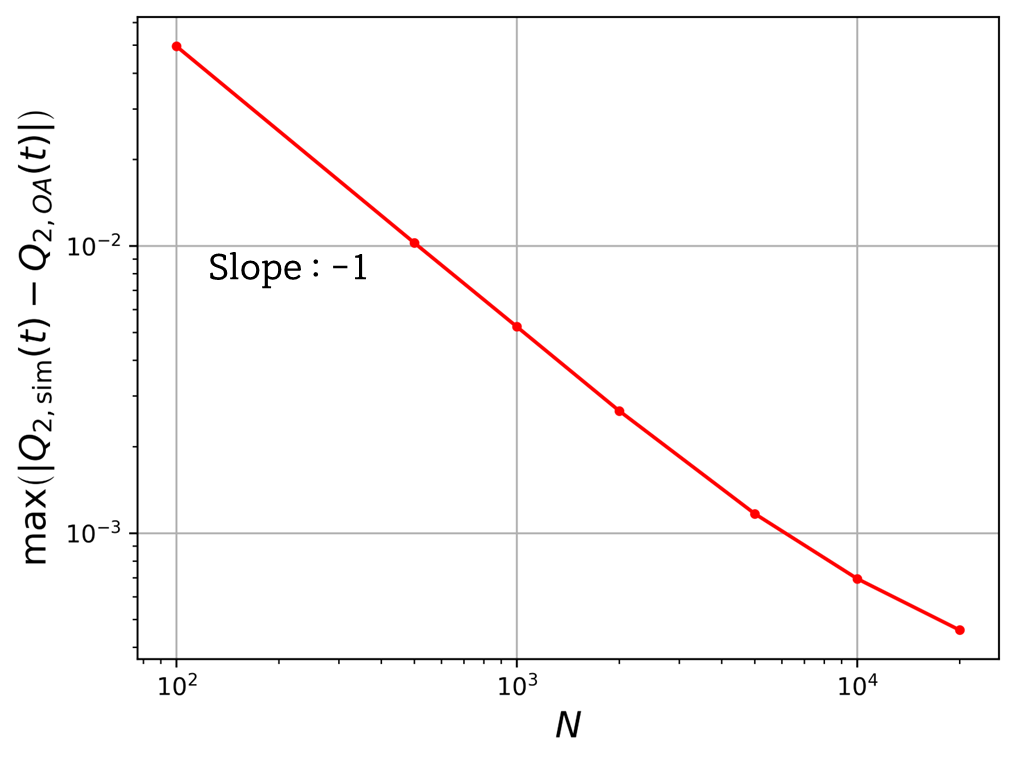}
    \caption{Maximum deviation of the Daido order parameters between simulation and theory versus system size $N$, confirming $O(N^{-1})$ scaling.}
    \label{fig:fig_finite_size_scaling}
\end{figure}

{The numerical simulation and the thermodynamic-limit OA prediction compared in Figs.~3(c) and 3(d) of the main text show a small residual discrepancy.}
To examine finite-size effects, we measured the maximum deviation between the microscopic simulation and the reduced OA prediction,
\begin{equation}
\Delta_N
=
\max_t
\left|
Q_{2,\mathrm{sim}}(t)-Q_{2,\mathrm{OA}}(t)
\right|.
\end{equation}
By increasing $N$, we find that the deviation decreases as
\begin{equation}
\Delta_N\sim O(N^{-1}).
\end{equation}
{As shown in Fig.~\ref{fig:fig_finite_size_scaling},} this scaling indicates that the residual discrepancy originates from finite-size fluctuations and vanishes in the thermodynamic limit.
    
\section{Dependence on initial conditions}

\subsection{Perturbations of the initial phases}

\begin{figure}[htbp]
\centering
\includegraphics[width=\linewidth]{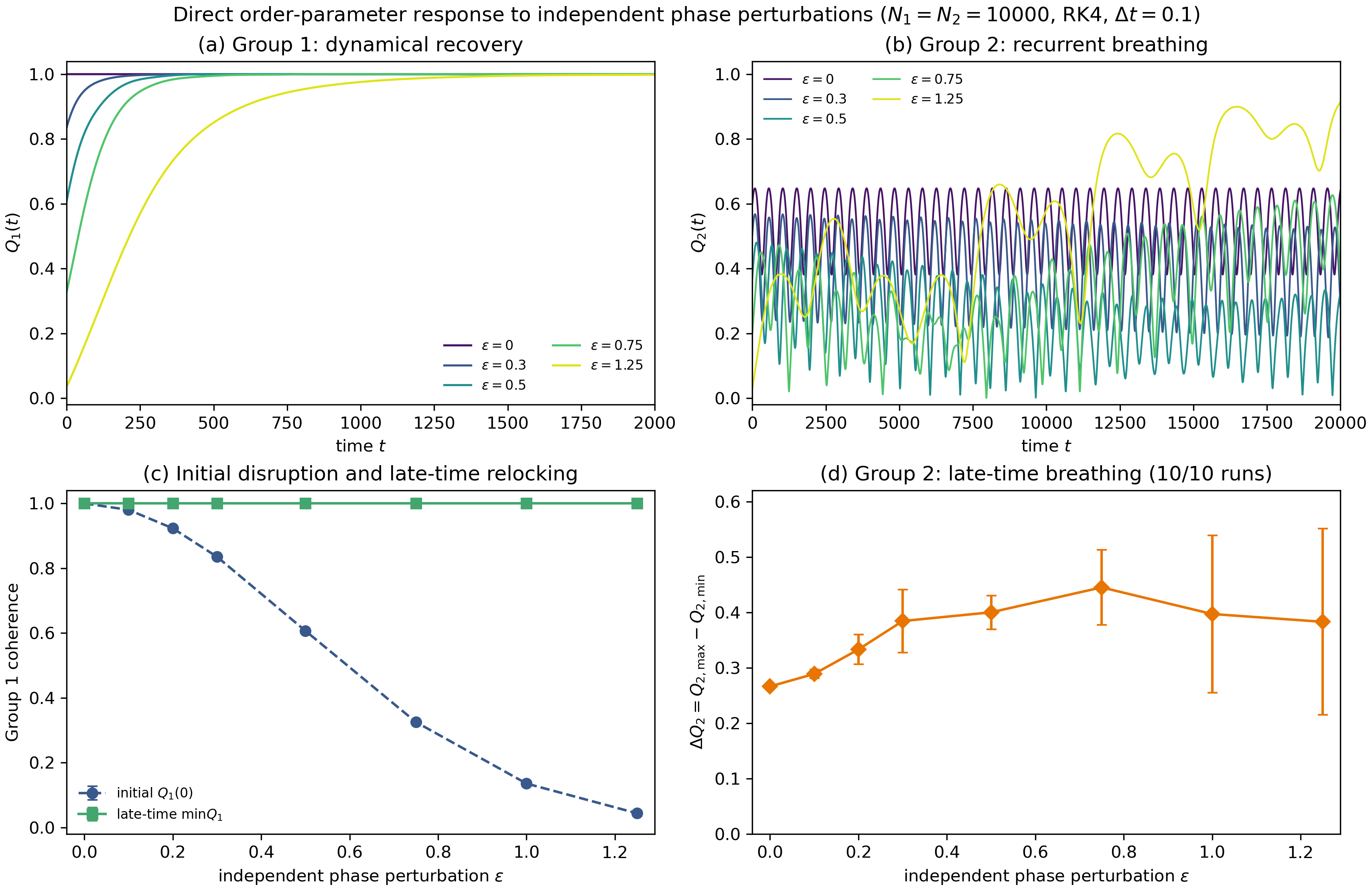}
\caption{
Response of the order parameters to independent perturbations of the
initial phases.
(a) Representative $Q_1(t)$ trajectories showing the recovery of
$Q_1\simeq1$.
(b) Corresponding $Q_2(t)$ trajectories showing persistent oscillatory
behavior after the perturbation.
(c) Initial $Q_1(0)$ and the minimum of $Q_1(t)$ over the late-time
observation window.
(d) Late-time breathing amplitude
$\Delta Q_2=Q_{2,\max}-Q_{2,\min}$.
We tested $\epsilon=0.1,0.2,0.3,0.5,0.75,1.0$, and $1.25$, using ten
independent realizations for each value; $\epsilon=0$ denotes the
unperturbed reference.
The late-time statistics were calculated over $3000\le t\le5000$ for
$\epsilon\le0.75$ and $10000\le t\le20000$ for $\epsilon=1.0$ and
$1.25$.
The unperturbed initial conditions and simulation parameters are
$Q_1(0)=1$, $Q_2(0)=0.60$, $\eta_1=0.55$, $\eta_2=0.50$,
$N_1=N_2=10000$, $\alpha=\pi/4$, $\mu=0.5$, and $\Delta t=0.1$.
Symbols and error bars indicate the mean and standard deviation,
respectively.
}\label{fig:initial_phase_perturbations}
\end{figure}

To test whether the breathing chimera state depends on the precisely
prepared initial phase distribution, we performed additional microscopic
simulations in which the initial phase of every oscillator in both groups
was independently perturbed according to
\begin{equation}
\theta_i^\sigma(0)
\rightarrow
\theta_i^\sigma(0)+\delta_i^\sigma,
\qquad
\delta_i^\sigma\sim\mathcal{N}(0,\epsilon^2).
\end{equation}
As shown in Fig.~\ref{fig:initial_phase_perturbations}, Group~1 recovers
$Q_1\simeq1$, while Group~2 retains persistent oscillations in $Q_2(t)$
for every tested perturbation strength and realization. Thus, the
breathing chimera state persists under the tested initial-phase
perturbations over the observation windows considered.

\subsection{Low initial coherence}

\begin{figure}[htbp]
\centering
\includegraphics[width=1\linewidth]{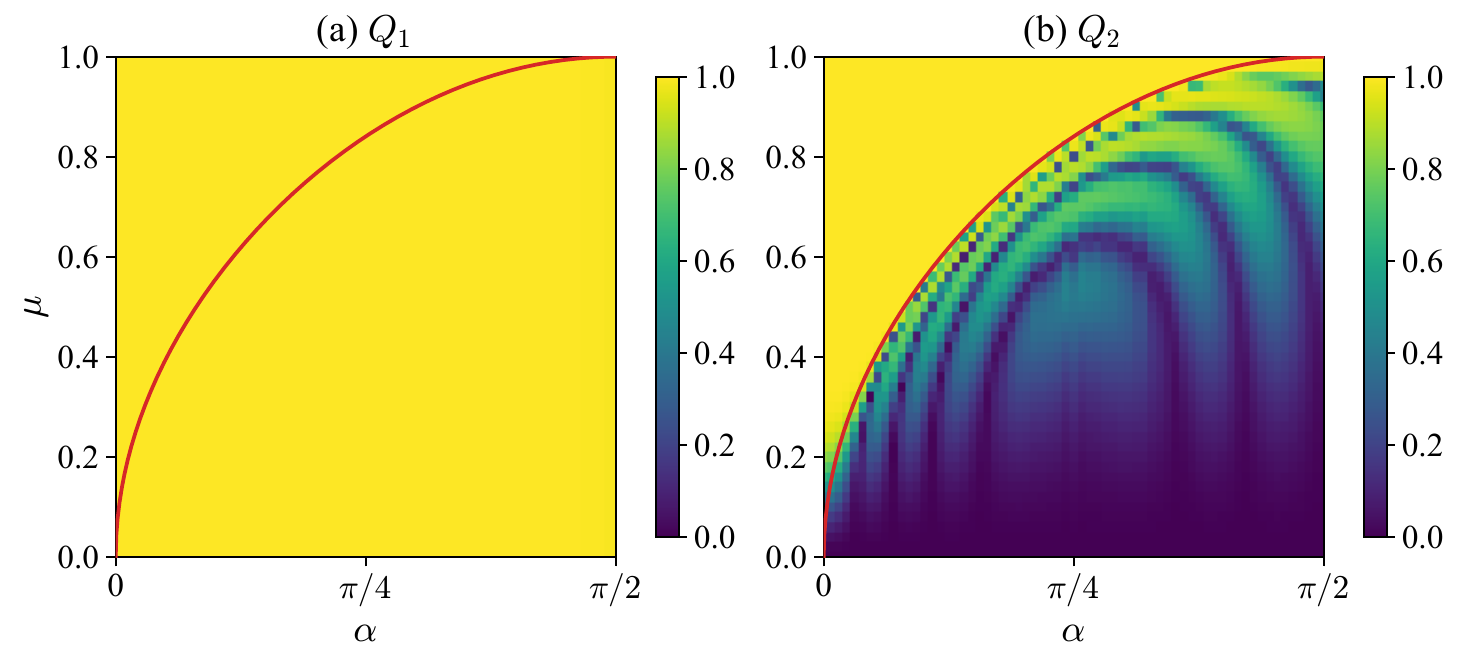}
\caption{
Coherence snapshots in the $(\alpha,\mu)$ space for the low-coherence initial condition
$Q_1(0)=0.1$ and $Q_2(0)=0$.
Panels (a) and (b) show $Q_1$ and $Q_2$, respectively, at $t=20000$.
The red curve denotes the theoretical boundary
$\mu=\sqrt{\sin\alpha}$.
The simulations use $N_1=N_2=10000$, $\eta_1=0.55$, and
$\eta_2=0.50$.
The corresponding temporal evolution is shown in Supplementary
Video~S3.
}
\label{fig:different_initial_coherence}
\end{figure}

To examine whether the long-time coherence and breathing dynamics are
directly imposed by the initial values of $Q_1$ and $Q_2$, we performed
additional simulations with the low-coherence initial condition
$Q_1(0)=0.1$ and $Q_2(0)=0$.
As shown in Fig.~\ref{fig:different_initial_coherence}, Group~1 develops
strong coherence in the breathing chimera region. Supplementary
Video~S3 further shows that Group~2 develops partial coherence with
persistent oscillations in $Q_2(t)$. These results show that the
long-time coherence and breathing behavior arise through the interaction
rather than being directly imposed by the initial values of $Q_1$ and
$Q_2$.

\section{Supplementary Video}
\noindent\textbf{Supplementary Video S1.}
Temporal evolution of the phase diagram in the $(\alpha,\mu)$ parameter space.
The left panel shows the thermodynamic-limit result obtained from the OA ansatz, while the right panel shows the corresponding numerical simulation result with $N_1=N_2=10000$.
The color represents the instantaneous value of the Daido order parameter $Q_2$, and the red solid line denotes the theoretical boundary $\mu=\sqrt{\sin\alpha}$.
The video demonstrates the quantitative agreement between the reduced theory and the finite-size simulations during the breathing dynamics.

\medskip
\noindent\textbf{Supplementary Video S2.}
Temporal evolution of the phase distributions shown in Fig.~2 of the main text. Panel (a) shows $Q_1$ and $Q_2$, while panels (b) and (c) show the oscillator phases represented by $\theta_i$ and $2\theta_i\bmod 2\pi$, respectively. Group~1 maintains a phase-locked bimodal distribution, while the phase distribution of Group~2 changes periodically as $Q_2$ oscillates. The parameters are the same as those in Fig.~2.

\medskip
\noindent\textbf{Supplementary Video S3.}
Temporal evolution of the phase diagram in the $(\alpha,\mu)$ parameter space for the initial condition $Q_1(0)=0.1$ and $Q_2(0)=0$.
The four panels show the instantaneous values of the Daido order parameters $Q_1$ and $Q_2$ and the Kuramoto order parameters $R_1$ and $R_2$, obtained from numerical simulations with $N_1=N_2=10000$, $\eta_1=0.55$, and $\eta_2=0.50$.
The red solid line denotes the theoretical boundary $\mu=\sqrt{\sin\alpha}$.
The video shows that Group~1 develops strong coherence with $Q_1\simeq1$, while $Q_2$ exhibits persistent oscillations in the breathing chimera region.
